\newcommand{\beq}{\begin{equation}}
\newcommand{\eeq}{\end{equation}}
\newcommand{\bea}{\begin{eqnarray}}
\newcommand{\eea}{\end{eqnarray}}
\newcommand{\bear}{\begin{eqnarray*}}
\newcommand{\eear}{\end{eqnarray*}}

\newcommand{\rf}[1]{(\ref{#1})}

\def\bra#1{\langle #1|}
\def\ket#1{|#1\rangle} 
\def\sstyle{\scriptstyle}
\documentclass[pre,twocolumn,showpacs,groupaddress,floatfix]{revtex4}

\usepackage{epsfig}
\usepackage{psfrag}
\usepackage{graphicx}

\begin{document}
\title
{The pair annihilation reaction $D + D \rightarrow \emptyset $ in disordered 
media 
and conformal invariance}
\author{F. C. \surname{Alcaraz}}
\email{alcaraz@if.sc.usp.br}
\affiliation{Instituto de F\'{\i}sica de S\~ao Carlos, Universidade de S\~ao 
Paulo, \\
Caixa Postal 369, 13560-590, S\~ao Carlos, S\~ao Paulo, Brazil. 
\vspace{0.1cm}}
\author{V. \surname{Rittenberg}}
\email{vladimir@th.physik.uni-bonn.de}
\affiliation{Physikalisches Institut, Bonn University, 53115 Bonn, Germany, \\ 
Department of Mathematics and Statistics, University
of Melbourne, Parkville,   Victoria 3010, Australia}

\date{\today}

\begin{abstract}
The raise and peel model describes the stochastic model of a fluctuating 
interface separating a substrate covered with clusters of matter of 
different sizes, and a rarefied gas of tiles. The stationary state is obtained
when  adsorption compensates the desorption of tiles. This model is 
generalized to an interface with defects ($D$). The defects are either 
adjacent or  separated by a  cluster. If a tile hits the end of a 
cluster with a defect nearby, the defect hops at the other end of the cluster 
changing its shape. If a tile hits two adjacent defects, 
the defect  annihilate and  are replaced 
by a small cluster. There are no defects in the stationary state. 
 This model can be seen as describing the reaction $D + D \rightarrow 
\emptyset $, in which 
the particles (defects) $D$ hop at long distances changing the medium and 
 annihilate. Between the 
hops the medium also changes (tiles hit clusters changing their shapes). 
Several properties of this model are presented and some exact results are 
obtained using the connection of our model with a conformal invariant 
quantum chain. 

\end{abstract}

\maketitle
\section{ Introduction}

 The raise and peel model \cite{GNPR, ALR} which is a stochastic model of a 
fluctuating interface is, to our knowledge, the first example of a 
stochastic model which has the space-time symmetry of conformal 
invariance. This implies that the dynamic critical exponent $z = 1$ 
 and certain scaling properties of various correlation functions are known. 
 This model 
was extended  in order to take into account sources at the boundaries 
\cite{DGP, PP, PAR} keeping conformal invariance. 
 In all these cases, the 
stationary states have magic combinatorial properties.

 In the present paper we describe another extension of the raise and peel 
model keeping conformal invariance (see also Appendix A and B)
 by introducing defects on the interface. 
These defects hop at long distances in a medium which is changed by the 
hops. Between the hops the medium also changes. Finally when two defects 
touch, they can annihilate. The stationary state is the one as the 
original raise and peel model with no defects.
 
The whole process can be seen as a reaction $D + D \rightarrow \emptyset $, 
where $D$ is a 
defect, taking place in a disordered unquenched medium.

 In Sec. II we describe the model. Like the raise and peel model 
\cite{GNPR}, the present model comes from considering the action of a 
Hamiltonian expressed in terms of Temperley-Lieb generators on a vector 
space which is a left ideal of the Temperley-Lieb algebra. The ideal can be 
mapped on graphs which constitute the configuration space of the model. We 
 shortly review in Appendix A  the mathematical background of the 
model and refer for details to Refs.~\cite{PP, PAR}.
 
In Sec. III, using Monte Carlo simulations, we describe the long range 
hopping of defects and give the L\'evy flights probability distribution.

 In Sec. IV, again using Monte Carlo simulations, starting with a 
configuration which consists only of defects, we study the variation in 
time $t$ of their density for a lattice of size $L$. We obtain the scaling 
function which gives the number of defects in terms of $t/L$ and show how 
conformal invariance gives some of its properties. In the thermodynamic 
limit the density decreases in time like $1/t$ as is expected since in a 
conformal invariant theory time and space are on equal footing. 
In Sec. V we present our conclusions.

\section{ The raise and peel model with defects.}\label{sect2}

We consider an interface of a one-dimensional lattice with  $L+1$ sites.
 An interface is formed by attaching at each site a non-negative 
integer height $h_i$ ($i = 0,1,\ldots,L$). We take $h_0 = h_L = 0$. If for two 
consecutive sites $j$ and $j+1$ we have $h_j = h_{j+1} = 0$, on the link 
connecting the two sites we put an arrow called defect (see Fig. 1). For 
the remaining sites, the heights obey the restrict solid-on-solid 
(RSOS) rules:
\begin{equation} \label{2.1}
h_{i+1} -h_i = \pm 1, \;\;\;\;\;\; h_i \geq 0.
\end{equation}

\begin{figure}[ht!]
\centering
{\includegraphics[angle=0,scale=0.40]{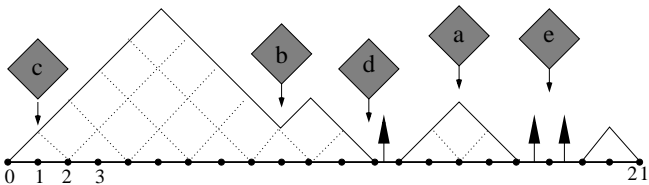}}
\caption{
 One of the 
  configurations for $L = 21$ (22 sites). There are 3 
defects 
(arrows on the links) and 3 clusters. Also shown are 5 tiles (tilted 
squares) (a)-(e)
belonging to the gas. When a tile hits the surface the effect is 
different in the five cases. }
\label{fig1}
\end{figure}

A domain in which the RSOS rules are obeyed 
$\{h_j=h_l=0, h_k >0, j<k<l\}$ is called a cluster. There are 
3 clusters and 3 defects in Fig.~1 ($L = 21$). There are ${L \choose [L/2]}$ 
possible
configurations of the interface (we denote by $[x]$ the integer part of $x$). 

There is a simple bijection between the configurations  
of interfaces and defects, where   Fig.~1 is an example, and ballot 
paths \cite{PP}. 
A ballot path is obtained if one follow the RSOS rules \rf{2.1}, take $h_L=0$ but
let $h_0$ free ($0\leq  h_L \leq L$).
This  fact was used in \cite{PAR} to define 
another stochastic model than the one described below.
 In the case $L = 4$, the six possible 
configurations are shown in Fig.~2. The configuration shown in Fig.~2b has 2 
defects and one cluster, while there are no defects in Fig.~2f. 
\begin{figure}[ht!]
\centering
{\includegraphics[angle=0,scale=0.38]{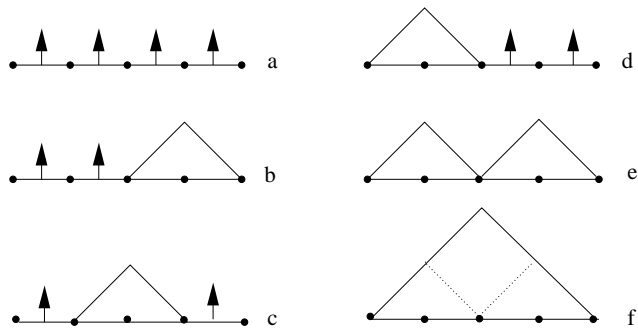}}
\caption{
 The six configurations for $L = 4$ (5 sites). In the stationary 
state only the RSOS configurations e) and f) occur.}
\label{fig2}
\end{figure}

 We consider the interface separating a film of tiles (clusters with 
defects) from a gas of tiles (tilted squares). 
 The evolution of the system (Monte Carlo steps) is given by the 
following rules. With a probability $P_i=\frac{1}{L-1}$ a tile from the gas 
hits site $i$
($i = 1,\ldots,L-1$). As a result of this hit, the following effects can 
take place:
\begin{itemize}

\item[a)] The tile hits a local maximum of a cluster ("a" in Fig.~1). 
The 
 tile is reflected.

\item[b)] The tile hits a local minimum of a cluster ("b" in Fig.~1). The tile 
is adsorbed.

\item[c)] The tile hits a cluster and the slope is positive 
($h_{i+1}>h_i>h_{i-1}$) ("c" in Fig.~1). The tile is reflected after 
triggering the 
desorption of a layer of tiles from the segment ($h_j>h_i=h_{i+b}$, $j = 
i+1,\ldots,i+b-1$), i.e.,  $h_j\rightarrow h_j -2,$ $j=i+1,...,i+b-1$. The 
layer contains $b 
- 1$ tiles (this is an odd 
number). Similarly, if the slope is negative ($h_{i+1}<h_i<h_{i-1}$, the 
tile is reflected after triggering the desorption of a layer of tiles 
belonging to the segment ($h_j> h_i = h_{i-b}$, $j+i-b+1,\ldots,i-1$).

\item[d)] The tile hits the right end of a cluster $h_j> h(i-c) = h(i) = 0$
($j =i-c+1,...,i-1$) and $h(i+1)=0$. The link ($i,i+1$) contains a defect 
("d" in Fig.~1). The defect hops on the link ($c,c+1$) after triggering 
the desorption of a layer of tiles ($h_j\rightarrow  h_j-2$, 
$j=i-c+1,...,i-1$)
 and 
the tile is adsorbed producing a new small cluster ($h_{i-1}=h_{i+1}=0, 
h_i=1$) (see Fig.~3). If the defect is at the left end of a cluster, the 
rules are similar, the defect hops to the right after peeling the 
cluster, and a new small cluster appears at the end of the old one.
\begin{figure}[ht!]
\centering
{\includegraphics[angle=0,scale=0.40]{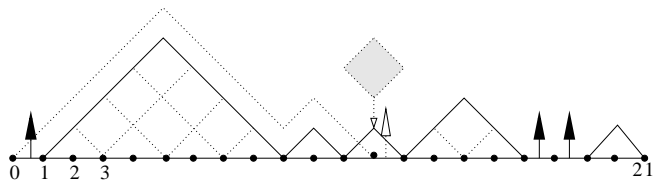}}
\caption{
 The new profile after the tile "d" in Fig.~1 has hit the 
surface at the right end of a cluster. The defect hops to the left end 
of the cluster, peeling one layer and a  new small cluster appears at the 
right of the old cluster.}
\label{fig3}
\end{figure}

\item[e)] The tile hits a site between two defects ($h_{i-1}=h_i=h_{i+1}=0$). 
This is the case "e" in Fig.~1. The two defects annihilate and in their 
place appears a small cluster ($h_{i-1}=h_{i+1}=0, h_i=1$). See Fig.~4.
\end{itemize}
\begin{figure}[ht!]
\centering
{\includegraphics[angle=0,scale=0.40]{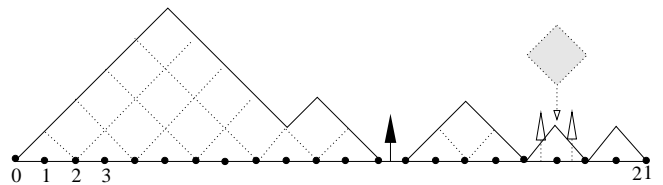}}
\caption{
 The new profile after the tile "e" has hit the surface between 
two defects. The defects have disappeared and in their place one gets 
a new small cluster.}
\label{fig4}
\end{figure}

 To sum up. The defects ($D$) hop non-locally in a disordered (not 
quenched) medium which changes between successive hops (local 
adsorption and nonlocal desorption take place in the clusters). During 
the hop the defect peels the cluster and therefore also changes the 
medium. The 
annihilation reaction $D + D \rightarrow \emptyset $  is local. If one starts 
the 
stochastic process with a certain configuration (for example, only 
defects as  in Fig.~\ref{fig3}a), due to the annihilation process, for $L$ 
even all the 
defects disappear and in the stationary state one has only clusters 
(RSOS configurations). The properties of the stationary states have 
been studied elsewhere \cite{GNPR, ALR}. In the case $L$ odd, in the 
stationary states one has one defect. In the next section we are going 
to see how this defect hops and will observe that the defect behaves 
like a random walker performing L\'evy flights. This will help us 
understand the annihilation process $D + D \rightarrow \emptyset $  described 
in Sec. IV.
 The rules described above  were  obtained by using a 
representation of the Temperley-Lieb algebra in a certain ideal \cite{DGP, 
PP, PAR}(see Appendix A). 
The finite-size scaling of the Hamiltonian eigenspectrum is 
known from conformal field theory (see Appendix B), therefore the 
physical properties of the model can be traced back to conformal 
invariance.

\section{ The random walk of a defect}

Before discussing the annihilation reaction of defects, it is useful to 
understand how defects hop. The simplest way to study the behavior of 
defects is to take the stationary states in the case $L$ odd when we 
have only a single defect. Although there is a lot of  information about these 
stationary states coming from combinatorics \cite{BGN,PAR} and Monte Carlo 
simulations \cite{PAR},  the results we present here are new.

 One asks what is the probability $P(s)$ for a defect to hop, at one 
Monte-Carlo step, at a distance $s$ (we assume $L$ very large). 
 We first see if on physical grounds, one can't guess the result. 
Let us assume that the defect behaves like a randon walker and that 
$P(s)$ describes L\'evy flights \cite{BG,HCF,WSUS,JMF}. This implies that for 
large values of $s$ we have:
\begin{equation} \label{3.1}
P(s) \sim \frac{1}{|s|^{1+\sigma}}.
\end{equation}
 If the random walker starts at a point $x = 0$ (for example in the 
middle of the lattice), at large values of $t$, the dispersion is \cite{JMF}:
\begin{equation} \label{3.2}
<x^2> \sim t^{\frac{2}{\sigma}}.
\end{equation}
 In a conformal invariant model, one has no other scales but the size 
of the system,  space and time are on equal footing therefore one 
has to have $\sigma = 1$.

 In Fig.~5 we show $P(s)$ as obtained from Monte Carlo simulations for 
systems of different sizes. One notices a data collapse for a large 
domain of $s$. A fit to the data for the largest lattice ($L = 4095$) 
gives, for large $s$:
\begin{equation} \label{3.3}
P(s) \approx \frac{2.25}{|s|^{2.06}},
\end{equation}
in  agreement with what we expected.

\begin{figure}[ht!]
\centering
{\includegraphics[angle=0,scale=0.46]{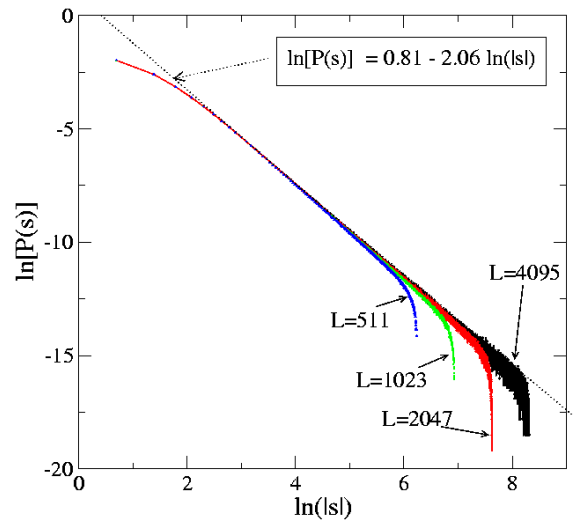}}
\caption{
 (Color online) The probability $P(s)$ for a defect to hop at a distance $s$ in 
units of lattice spacing. Monte Carlo simulations were done on systems 
of different sizes.}
\label{fig5}
\end{figure}

\section{ The density of defects at large times}

 We are now going to study the number of defects $N_d(t,L)$ as a function of 
time and lattice size taking  at $t = 0$ the configuration  where the 
lattice is covered by defects only 
(like in Fig.~2a).  An 
interesting and novel aspect of this study is the role of conformal 
invariance.
 Since there are no other scales in the system except $L$, we expect for 
large values of $t$ and $L$:
\begin{equation} \label{4.1}
N_d(t,L) = f(\frac{t}{L}).
\end{equation}

 In Fig.~6, we show $N_d(t,L)$  for various lattice sizes ($L$  
odd). One sees a nice data collapse except for very  small values of 
$t/L$ where the convergence is slower. A similar (but not identical!) 
function is obtained for $L$ even.
\begin{figure}[ht!]
\centering
{\includegraphics[angle=0,scale=0.46]{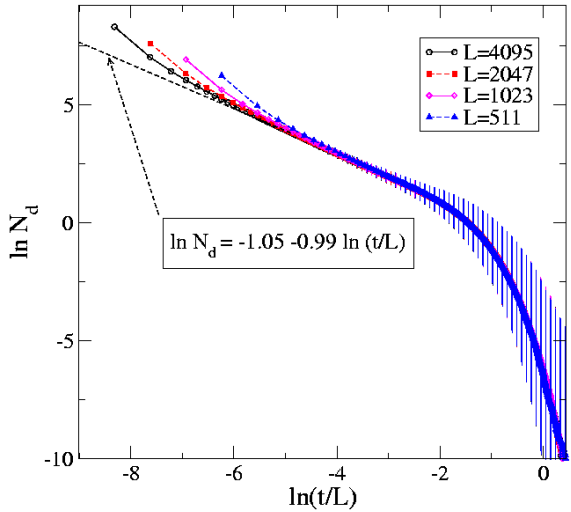}}
\caption{
 (Color online) The number of defects $N_d(t,L)$ as a function of $t/L$ for several lattice 
sizes ($L$ odd). At $t= 0$, $N_d(t=0,L) = L$. The error bars are also shown. 
The fitted linear curve shows that the 
density decreases as the inverse of time.}
\label{fig6}
\end{figure}

 We firstly discuss the behavior of $N_d$ for large values of $t/L$ (see 
Fig.~7). A fit to the data gives ($L$ odd):
\begin{equation} \label{4.2}
N_d(\frac{t}{L}) = A_1^{(o)} e^{-\lambda_1^{(o)}\frac{t}{L}} + 
A_2^{(o)} e^{-\lambda_2^{(o)}\frac{t}{L}} + \cdots ,
\end{equation}
where
\bea \label{4.3}
A_1^{(o)} &=& 6.75, \;\;\;\; A_2^{(o)} = 17.27, \nonumber \\
\lambda_1^{(o)} &=& 8.21, \;\;\;\; \lambda_2^{(o)} = 26.48.
\eea
We can now compare the data obtained from the fit with the finite-size 
scaling spectrum of the Hamiltonian (see Appendix B, Eqs. \rf{A10},\rf{A11}):
\bea \label{4.5}
\lambda_1^{(o)} &=& \frac{3\pi\sqrt{3}}{2} = 8.1620971\cdots, \nonumber \\
\lambda_2^{(o)} &=& \frac{3\pi\sqrt{3}}{2} \frac{10}{3}  = 27.20699\cdots. 
\eea
No prediction can be made about $A_1^{(o)}$ or $A_2^{(o)}$ since they are not 
universal, they depend on the  initial conditions. Notice that $A_2^{(o)} > 
A_1^{(o)}$ 
as it should be since the expansion should diverge for short times where 
we expect
\begin{equation} \label{4.6}
N_d \sim \frac{L}{t}
\end{equation}

\begin{figure}[ht!]
\centering
{\includegraphics[angle=0,scale=0.46]{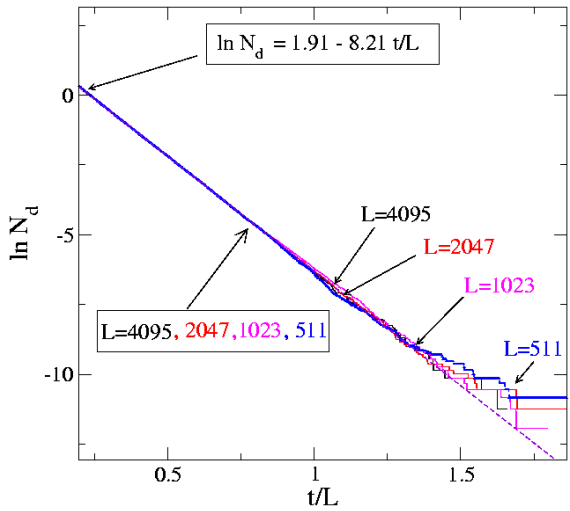}}
\caption{(Color online)
 The number of defects $N_d$ as a function of $t/L$ as in Fig.~6 
zoomed on the large time domain. The error bars, given in Fig.~\ref{fig6}, are not shown. }
\label{fig7}
\end{figure}

 A similar fit, done for $L$ even (the data are shown in Fig.~8), gives:
\begin{equation} \label{4.7}
N_d(\frac{t}{L}) = A_1^{e)} e^{-\lambda_1^{(e)}\frac{t}{L}} + 
A_2^{(e)} e^{-\lambda_2^{(e)}\frac{t}{L}} + \cdots ,
\end{equation}
with 
\bea \label{4.8}
A_1^{(e)} &=& 2.83, \;\;\;\; A_2^{(e)} = 6.93,\nonumber \\ 
\lambda_1^{(e)} &=& 2.71, \;\;\;\; \lambda_2^{(e)} = 16.64.
\eea
 We can again use the predictions of conformal invariance (see \rf{A10} and 
\rf{A11}) and get
\bea \label{4.10}
\lambda_1^{(e)} &=& \frac{3\pi\sqrt{3}}{2}\frac{1}{3} = 2.72069\cdots,\nonumber \\
\lambda_2^{(e)} &=& \frac{3\pi\sqrt{3}}{2}2 = 16.324194\cdots. 
\eea
to be compared with \rf{4.8}.
\begin{figure}[ht!]
\centering
{\includegraphics[angle=0,scale=0.46]{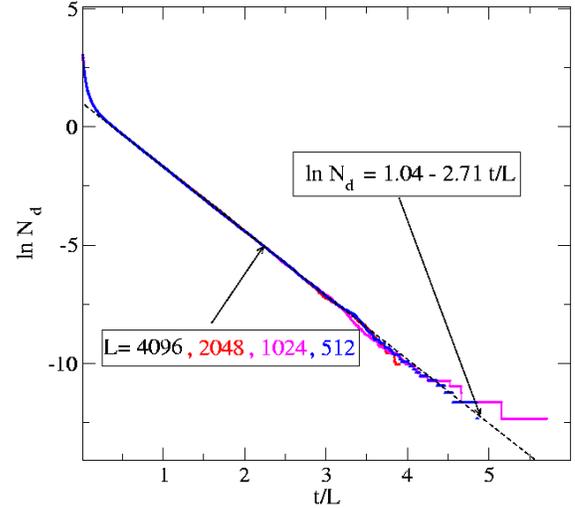}}
\caption{(Color online) 
 $N_d$ as a function of $t/L$ for large times for different lattice 
sizes ($L$ even). The error bars are of the same order as in Fig.~\ref{fig6}.}
\label{fig8}
\end{figure}

 In the small $t/L$ domain we get for $L$ even and odd:
\begin{equation} \label{4.11}
\rho = \frac{N_d}{L} \approx \frac{0.322}{t}.
\end{equation}
 In order to find the correction term in \rf{4.11}, we have computed 
$N_dt^2/L$ 
as shown in Fig.~9. We have obtained a straight line from which we get:
\begin{equation} \label{4.12}
\rho = \frac{0.322}{t} + \frac{0.334}{t^2} + \cdots .
\end{equation}
This last result  is the same for $L$ even and odd. Notice that the 
correction term in \rf{4.12} is not given by the scaling function \rf{4.1}.
\begin{figure}[ht!]
\centering
{\includegraphics[angle=0,scale=0.46]{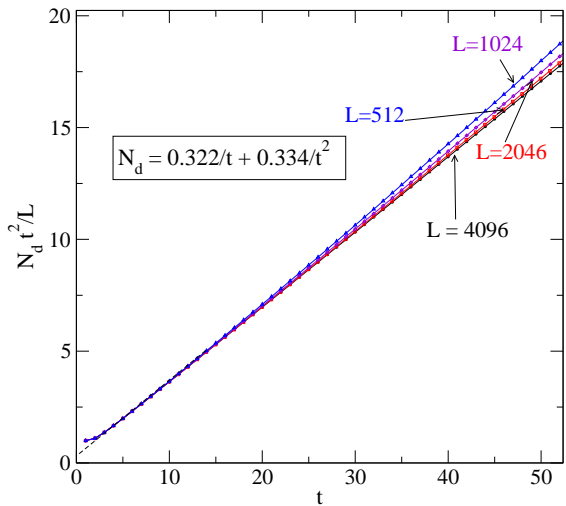}}
\caption{(Color online) 
 The density of defects times $t^2$ for short times. A linear fit to 
the data obtained for the largest lattice ($L = 4096$) gives \rf{4.12}.}
\label{fig9}
\end{figure}

 We have also computed the fluctuation of the density as a function of 
time and got:
\begin{equation} \label{4.13}
\frac{<\rho^2> - <\rho>^2}{<\rho>^2} \approx \frac{0.237}{t^{1.00}} .
\end{equation}

 We would like to compare our results with known results obtained for 
diffusion and annihilation reactions ( $ A + A \rightarrow \emptyset $ ) 
with L\'evy flights \cite{ALB,HH,VH,DV,THVL}. In one dimension, for L\'evy 
flights 
given by Eq.~\rf{3.1}, one gets \cite{HH}:
\begin{equation} \label{4.14}
\rho \sim  \left\{\begin{array}{rc}
t^{-\frac{1}{\sigma}}&\mbox{for}\quad \sigma>1 \\
\frac{\ln t}{t} &\mbox{for}\quad \sigma=1 \\
t^{-1}&\mbox{for}\quad \sigma<1 
\end{array} \right.
\end{equation}
the critical dimension being $d_c= \sigma$.

 If one compares \rf{4.14} for $\sigma = 1$,  
as obtained in Sec. III and \rf{4.12}, one notices the absence of the 
$\ln t$ 
correction. Such a term, if present, could   have been seen in our simulations 
(one 
observes that for large lattices, $\rho t$ converges to the value 
0.322 from 
above). Logarithmic corrections can also appear  in a conformal field 
theory if one has Jordan cells \cite{jordan} but there are no Jordan cells in 
the 
Hamiltonian \rf{A2} given in Appendix A \cite{PPe}.
 We believe that the discrepancy between the results of our model and 
those obtained for the reaction $A + A \rightarrow \emptyset $ 
comes from the fact that the two 
models have little in common.

\section{ Conclusions}

 We have presented an extension of the raise and peel model taking into 
account defects. The main property of this model is that conformal 
invariance is preserved. The model mimics a system in which particles move 
in a disordered unquenched medium doing L\'evy flights and changing the 
medium during the flights. Upon contact the defects annihilate. The 
properties of the system are simple and could be guessed on simple 
grounds based on conformal invariance. 
Conformal field theory enters in the description of the scaling 
function $N_d = f(t/L)$ ($N_d$ is the number of defects, $L$ the size of the 
system and $t$ is the time).

 The original raise and peel model \cite{ALR} (this is the present model with 
the 
defects absent) depends on a parameter $w$ which is the ratio of the 
desorption and adsorption rates. If $w = 1$, one has conformal invariance 
and the dynamic critical exponent $z = 1$. If one takes $0 < w < 1$, in the 
disordered medium one has less clusters and $z$ 
varies continuously  in the interval $0 < z < 1$. 
One can add defects to the model and 
repeat the exercise done in this paper for all values of $w$. In this 
case one expects to find defects making L\'evy flights with a probability 
distribution function:
\begin{equation} \label{5.1}
  P(s)\sim \frac{1}{ s^{1+z}}.
\end{equation}

\section*{Acknowledgments}
We thank H. Hinrichsen for a carefull reading of the manuscript.
This work has been partially supported by the Brazilian agencies FAPESP and 
CNPq, and
by  The Australian Research Council.


\newpage
\appendix

\section{ The connection between Temperley-Lieb stochastic processes
and the raise and peel model}

We shortly review this connection, for details see \cite{PPN,PP} and 
\cite{PAR}. 

Consider the Temperley-Lieb semigroup algebra $G$ defined by $L - 1$
 generators 
$e_j$ ($j=1,\ldots,L-1$) and the relations:

\bea
e_j^2 &=& e_j,\qquad e_je_{j\pm1}e_j = e_j,\nonumber \\
  e_je_k &=& e_ke_j \quad
{\rm for} \quad |j-k| > 1,
\label{B.1}
\eea
and the Hamiltonian 
\beq
                   H = \sum_{j=1}^{L-1} (1-e_j).
\label{B.2}
\eeq
In the basis $\{w_c\}$ of the words of $G$ 
(the regular representation of $G$), $H$ 
is a matrix satisfying $H_{a,b}\leq 0$ for 
$a \neq  b$ and $\sum_bH_{a,b}=0$. Such a matrix is an intensity 
matrix and defines a 
Markov process in continuum time given by the master equation
\begin{equation}\label{B.3}
\frac{d}{dt} P_a(t) = -\sum_b H_{a,b} P_b(t),
\end{equation} 
where $P_a(t)$ is the (unnormalized) probability to find the system in the 
state $\ket a$ at time $t$ and the rate for the transition $\ket b \to \ket a$  is given by 
$- H_{a,b}$, which is non-negative. The Hamiltonian \rf{B.2} has an eigenvalue 
equal to zero. The corresponding left eigenvector $\bra0$ is trivial, the 
right eigenvector $\ket0$ gives the probabilities in the stationary state:
\bea \label{B.4}
\bra0 H &=& 0, \;\;\; \bra0 = \sum_a \bra a, \nonumber \\
H \ket0  &=& 0, \;\;\; \ket0 = \sum_a P_a \ket a, \;\; P_a = \lim_{t \to \infty}  P_a(t).
\eea

 $H$ defined by \rf{B.2} gives a Markov process not only if it acts in the 
vector space of the regular representation but also if it acts in the 
vector space of a left ideal $I$ because the generators $e_j$ map the left 
ideal into the left ideal:
\begin{equation}\label{B.5}
e_j I= I.    
\end{equation}

 An easy way to define the left ideal in which we are interested and the 
action of the generators on this ideal is to use the language of graphs.

 The generators $e_j$ can be pictorially represented by 
\beq
e_j \quad = \quad
\begin{picture}(140,20)
\put(2,-10){\epsfxsize=135pt\epsfbox{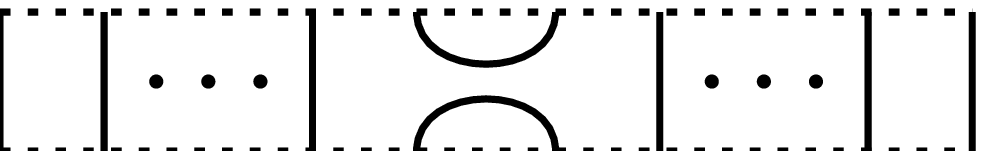}}
\put(.5,-18){$\sstyle 1$}
\put(15,-18){$\sstyle 2$}
\put(38,-18){$\sstyle j-1$}
\put(58,-18){$\sstyle j$}
\put(72,-18){$\sstyle j+1$}
\put(89,-18){$\sstyle j+2$}
\put(112,-18){$\sstyle L-1$}
\put(135.5,-18){$\sstyle L$}
\end{picture}
\hspace{10pt}\vspace{28pt}
\label{B.6}
\eeq
\vspace{1cm}

\begin{figure}[ht!]
\centering
{\includegraphics[angle=0,scale=0.30]{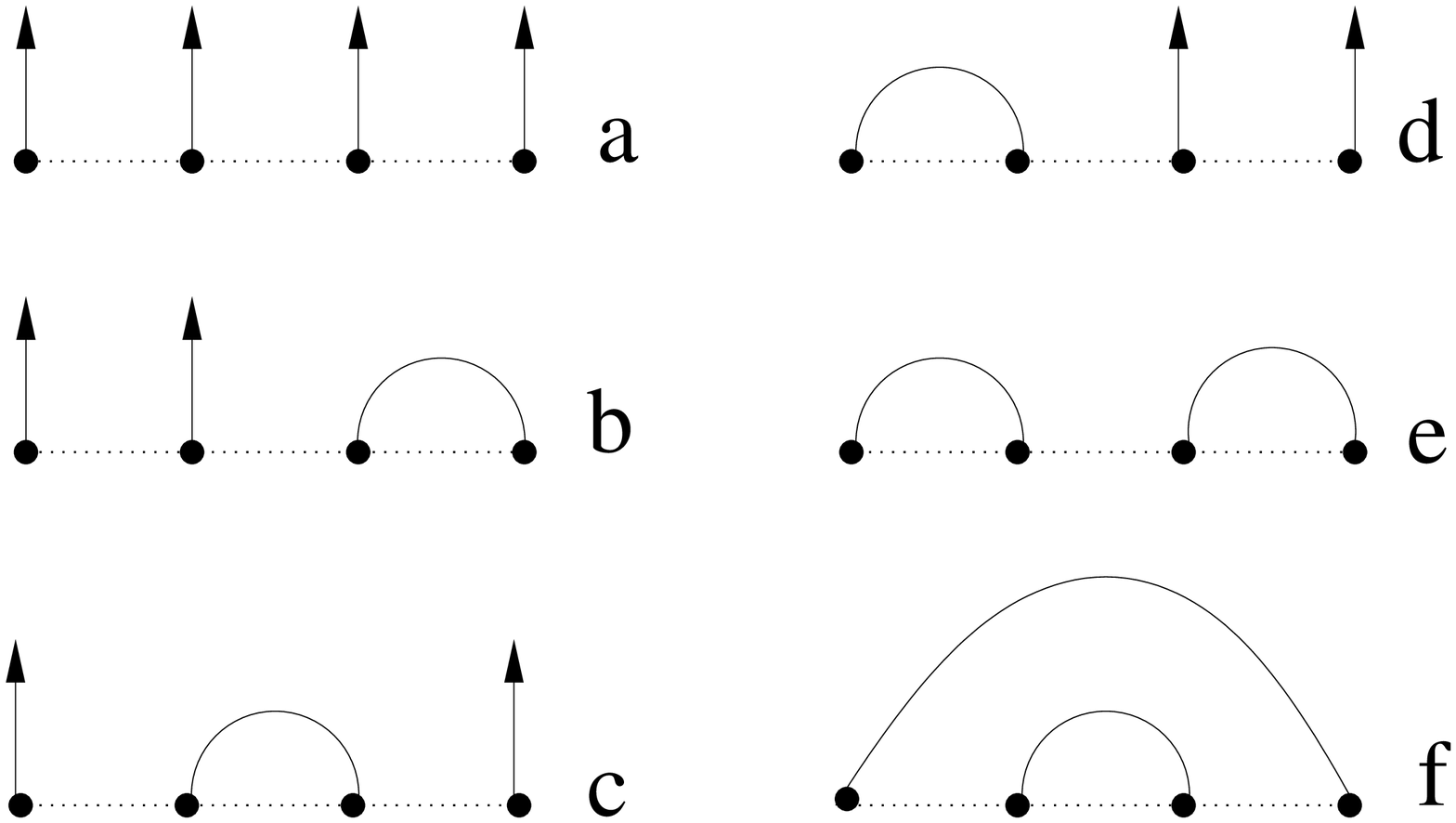}}
\caption{
 The six links-defects diagrams for $L=4$. The diagrams a-f 
correspond to the a-f RSOS configurations of Fig.~\ref{fig2}.}
\label{fig10}
\end{figure}

 The elements of the ideal can be represented by links-defects 
diagrams. They can be obtained in the following way (see Fig.~\ref{fig10} for $L = 4$): 
take $L$ sites. If a site is not connected to another one, draw a vertical 
arrow. Two sites can be connected by a link. The links don't cross each 
other and the arrows can't cross the links. For a given $L$ the number of 
diagrams with $m$ defects is:
\begin{equation}\label{B.7}
C_{L,m} ={L \choose [ \frac{L-m+1}{2}]} - {L \choose [\frac{L-m-1}{2}]}
\end{equation}
and the total number of diagrams is
\begin{equation}\label{B.8}
\sum_{s=0}^{L/2} C_{L,2s+(L \;\; \mbox{\scriptsize mod} 2)} = {L \choose [\frac{L}{2}]},
\end{equation}
where $[x]$ is the integer part of $x$.

\begin{figure}[ht!]
\centering
{\includegraphics[angle=0,scale=0.30]{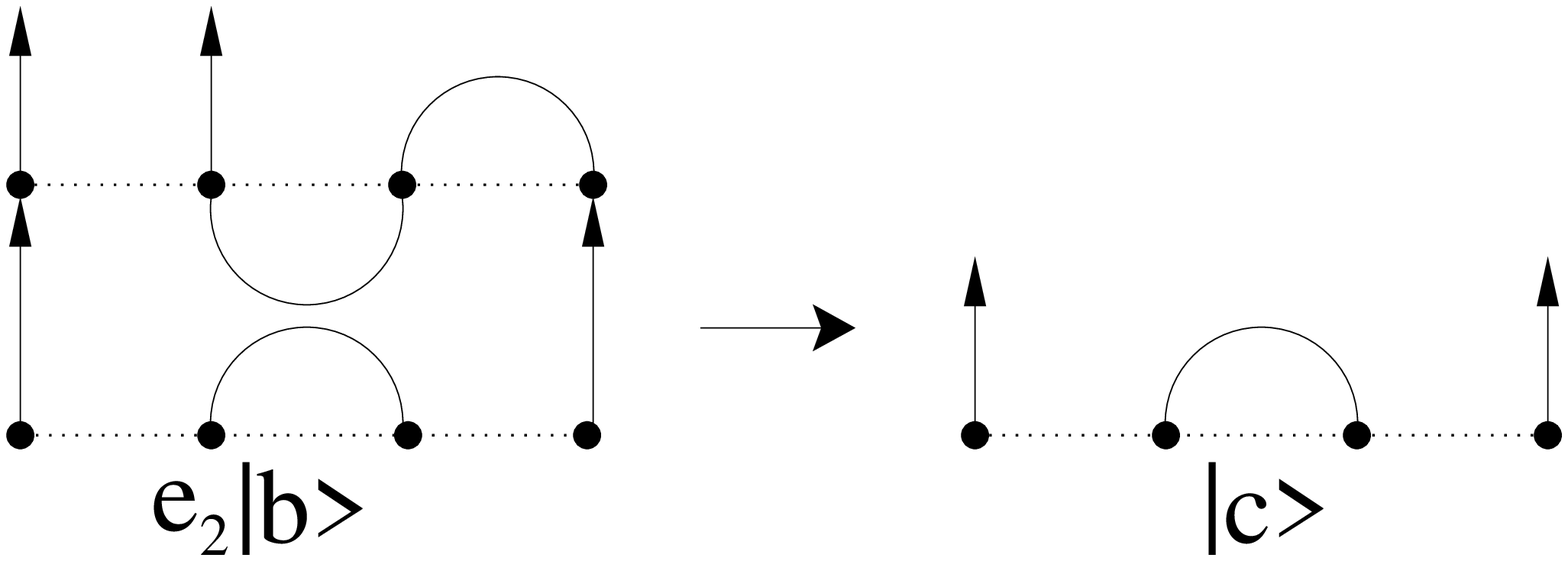}}
\caption{
 The action of $e_2$ on the diagram $b$ of Fig.~\ref{fig10}.}
\label{fig11}
\end{figure}

 The action of $e_j$ on a links-defects diagram is given by placing the 
graph of $e_j$ underneath the first diagram, removing the closed loops and 
the intermediate dashed line. Next one contracts the links in the  
composite picture. In Fig.~\ref{fig11} we show the action of $e_2$ on the diagram $b$ of 
figure \ref{fig10}.

 
The action of the Hamiltonian \rf{B.2} in the vector space given in Fig.~\ref{fig10}  
is
\beq \label{B.9}
H =  \left( \begin{array}{@{}cccccc@{}}
3  & 0  & 0 & 0 & 0 & 0 \\
-1 & 2 & -1 & 0 & 0 & 0\\
-1 & -1 & 2 & -1 & 0 & 0\\
-1 & 0 & -1 & 2 & 0 & 0 \\
0 & -1 & 0 & -1 & 1 & -2 \\
0 & 0  & 0 & 0  & -1 & 2 
\end{array}\right).
\eeq
Notice that $H$ has a block triangular form. The stationary state $\ket0  = 2\ket e  
+ \ket f$ contains only the two states without arrows (defects) $\ket e$ 
 and $\ket f$. 
The various transition rates can be obtained from the matrix elements of $H$.

 In Appendix B we are going to use a $2^L$ dimensional representation of the 
$L - 1$ generators $e_j$ and of the Hamiltonian \rf{B.2}. In this representation, 
the Hamiltonian describes a spin $1/2$ quantum chain. Where can we find the 
eigenvalues of the left ideal (their number is given by \rf{B.8}), among the $2^L$ 
eigenvalues of the quantum chain? We are going to give an "almost 
correct" explanation. We take again the case $L = 4$ as an 
example. If on each of the $4$ sites of the chain one takes a spin $1/2$ 
representation of $sl(2)$, one finds the representation with spin 0 (2 
times), spin 1 (3 times) and spin 2 (one time). If for each representation 
containing $2s+1$ states ($s$ is the spin) one takes only the highest weight 
states, one gets precisely 6 states (the vertical arrows in Fig.~\ref{fig10}  
corresponding to up spins).

 We give now the correspondence between the links-defects diagrams and the 
RSOS configurations considered in section 2.
 For  a links-defects diagram with $L$ sites ($i = 1,2,\ldots,L$) take the dual 
lattice with $L+1$ sites (on each bond between the sites $i$ and $i + 1$ of the 
links-defects diagram you take the  site $i$ on the dual lattice. On the 
dual lattice we have the sites $j$ ($j = 0,1,\ldots,L$). An arrow (defect) on the 
site $i$ on the links-defects diagram, stays unchanged on the dual lattice (it is 
on the bonds of the dual lattice). For the links, one proceeds as follows: 
one takes a site on the dual lattice and a vertical line on this site. One 
counts how many links are cut by the vertical line and one takes a vertex     
with a height $h$ equal to the number of intersections. Figs.~\ref{fig2} and \ref{fig10}  
illustrate the rules.

\section{ The finite-size scaling limit of the Hamiltonian eigenspectrum.  
 Results from conformal field theory.}

We are going to give a brief description of the time evolution operator of 
the stochastic model described in Sec. 2.
 Firstly we consider the spin $\frac{1}{2}$ quantum chain defined by the 
Hamiltonian 
\begin{equation} \label{A1}
H = \sum_{i=1}^{L-1} (1-e_i), 
\end{equation}
where 
\begin{equation}\label{A2}
e_i = \frac{1}{2}[ 
\sigma_i^x\sigma_{i+1}^x 
+ \sigma_i^y\sigma_{i+1}^y 
-\frac{1}{2} \sigma_i^z\sigma_{i+1}^z
+i\frac{\sqrt{3}}{2}(\sigma_i^z - \sigma_{i+1}^z)],
\end{equation}
and $\sigma^x$, $\sigma^y$, $\sigma^z$ are Pauli matrices. The Hamiltonian 
\rf{A1} 
commutes with
\begin{equation} \label{A3}
S^z = \frac{1}{2}\sum_{i=1}^L \sigma_i^z.
\end{equation}

 In the continuous time limit, the evolution of the system is given by a 
Hamiltonian $H^e$ which corresponds to the 
subspace of highest weight $U_q(Sl(2)$ representations ($q=\exp{i\pi/3}$) \cite{pasquier-saleur}.
 There are ${L \choose [L/2]}$
states in these two sectors ($[x]$ is the integer part of $x$). If we denote 
by $E_r$ ($r = 0,1,\ldots$) the energy levels in non decreasing  order: 
$E_0 =0 < E_1 \leq E_2 \leq \cdots$, the
  partition function giving the finite-size scaling limit of the 
spectrum of $H^e$ is defined as follows:
\begin{equation} \label{A6}
Z(q) = \lim_{L \to \infty} Z_L(q) = \lim_{L\to \infty} \sum_n q^{LE_n/\pi v_s},
\end{equation}
where $v_s = 3\sqrt{3}/2$. One can show \cite{BS} that $Z(q)$ has the 
expression
\begin{equation} \label{A7}
Z(q) = \sum_s \zeta_s(q).
\end{equation}
Here $s$ is the spin, taking the values   $s = 0,1,2,\ldots $ for $L$ even and 
$s = \frac{1}{2},\frac{3}{2}, 
\frac{5}{2}$ for $L$ 
odd, and 
\begin{equation} \label{A8}
\zeta_s(q) = q^{\Delta_s} (1- q^{2s+1}) \prod_{n=1}^{\infty}(1-q^n)^{-1},
\end{equation}
where 
\begin{equation} \label{A9}
\Delta_s = \frac{s(2s-1)}{3}, 
\end{equation}

 Moreover, for large lattice sizes,  the energies are (see \rf{A6} and 
\rf{A8})
\begin{equation} \label{A10}
E = \frac{3\pi \sqrt{3}}{2L}(\Delta_s+k),
\end{equation}
where $k$ is an integer.

 The Hamiltonian $H^e$ has a block diagonal form. The states with no defects 
($L$ even) and those with one defect ($L$ odd) are in one block. This is the 
$s = 0$
($s = \frac{1}{2}$) part of \rf{A7}. The states with defects ($L$ even) and 
more 
than one defect ($L$ odd) correspond to higher spins. In Sec. 3 we found 
that the following values of $\Delta_s$ were useful:
\begin{eqnarray} \label{A11}
\Delta_1 &=& \frac{1}{3}, \; \Delta_2 = 2 \;\;\;\;\;(L \;\;\mbox{even}) 
\nonumber \\
\Delta_{\frac{3}{2}} &=& 1, \; \Delta_{\frac{5}{2}} = \frac{10}{3}\;\;\;\;\; 
(L\;\; \mbox{odd})  
\end{eqnarray}

\begin{thebibliography}{99}
\bibitem{GNPR}  J. de Gier, B. Nienhuis, P. Pearce and V. Rittenberg , J. 
Stat. Phys. {\bf 114} 
(2004) 1.
\bibitem{ALR} F. C. Alcaraz, E. Levine and V. Rittenberg, J. Stat. Mech (2006) 
P08003.
\bibitem{DGP} J. de Gier and P. Pyatov, J. Stat. Mech. (2004) P0403002.
\bibitem{PP} P. Pyatov, J. Stat. Mech. (2004) P09003.
\bibitem{PAR} F. C. Alcaraz, P. Pyatov and V. Rittenberg, to be published.
\bibitem{BGN} M. T. Batchelor, J. de Gier and B. Nienhuis, J. Ph
ys. A {\bf 34}
(2001) L265.
\bibitem{BG}  J-P. Bouchaud and A. Georges, Phys. Rep. {\bf 195} (1990) 127.
\bibitem{HCF} H. C. Fogedby, Phys. Rev. Letters {\bf 73} (1994) 2517. 
\bibitem{WSUS} E. R. Weeks, T. H. Solomon, J. S. Urbach and H. L. Swinney in 
"L\'evy Flights
and Related Topics in Physics", edited by M. F. Schlessinger, G. M. Zaslavsky
and U. Frish (Springer-Verlag, Heidelberg, 1995).
\bibitem{JMF} S. Jespersen, R. Melzer and H. C. Fogedby, Phys. Rev. E {\bf 59} 
(1999) 2736.
\bibitem{ALB} E. V. Albano, J. Phys. A {\bf 24} (1991) 3351.
\bibitem{HH} H. Hinrichsen and M. Howard Eur. Phys. J. B {\bf 7} (1999) 635.
\bibitem{VH}  D. Vernon and M. Howard, Phys. Rev. E {\bf 63} (2001) 041116,  
cond-mat/0011475.
\bibitem{DV} D. Vernon, Phys. Rev. E {\bf 68} (2003) 041103,  
cond-mat/0304376.
\bibitem{THVL}  U. Tauber, M. Howard and B. Vollmayr-Lee, J. Phys. A {\bf 38} 
(2005) R79.
\bibitem{jordan} M. Gaberdiel, Int. J. Mod. Phys. A {\bf 18} (2003) 4593.
(2003) R35.
\bibitem{PPe} Paul Pearce, private communication.
\bibitem{PPN} P. P. Pearce, V. Rittenberg, J. de Gier and B. Nienhuis, J. Phys. A {\bf 35} (2002) L661.
\bibitem{pasquier-saleur} V. Pasquier and H. Saleur, Nucl. Phys. B 
{\bf 330} (1990) 523.
\bibitem{BS}  H. Saleur and M. Bauer , Nucl. Phys. B {\bf 320} (1989) 591.
{\bf 10} 


\end{thebibliography}
\end{document}